\documentstyle[amssymb,11pt]{article}

\begin{document}

\title{Corrections to the Newton and Coulomb potentials caused by effects of
spacetime foam}
\author{A.A. Kirillov \\
{\em Institute for Applied Mathematics and Cybernetics} \\
{\em 10 Uljanova Str., Nizhny Novgorod, 603005, Russia}\\
e-mail: kirillov@unn.ac.ru \and D. Turaev \\
{\em Weierstrass Institute for Applied Analysis and Stochastisity} \\
{\em Mohrenstrasse 39, D-10117, Berlin, Germany} \\
e-mail: turaev@wias-berlin.de}
\date{}
\maketitle

\begin{abstract}
We use the extended quantum field theory (EQFT) \cite{k99} to explore possible 
observational effects of the spacetime. It is shown that as it was expected 
the spacetime foam can provide quantum bose fields with a cutoff at very small 
scales, if the energy of zero - point fluctuations of fields is taken 
into account. It is also shown that EQFT changes the behaviour of massless 
fields at very large scales (in the classical region). We
show that as $r\gg 1/\mu $ the Coulomb and Newton forces acquire the
behaviour $\sim 1/r$ (instead of $1/r^{2}$).
\end{abstract}

It is commonly believed that quantum gravity effects (spacetime foam) should
provide a cutoff for quantum field theory. An extension of the quantum field
theory (EQFT) which allows to account for the spacetime foam effects was
suggested by one of us in Ref. \cite{k99}. This extension involves a
variable number of Bose fields and introduces a new operator $N\left(
k\right) $ which is the density of the number of fields in the space of
modes. The standard QFT represents a particular sector of the extended
theory when $N\left( k\right) =1$. In this letter we show that this operator
causes a modification of Green functions $\widetilde{D}\left( k\right)
=D\left( k\right) N\left( k\right) $ in all internal lines of the diagram
technique. At very small scales $k\ll k^{\ast }$ it indeed can provide a
cutoff, if we take into account the energy of zero-point fluctuations of
fields. From the other side, in the case of massless fields this operator
causes essential modification at large scales $k\ll \mu $ ($\mu $ is the
Fermi energy, e.g., see Ref. \cite{k99}).

First we note, that from a phenomenological view point the operator $N\left(
k\right) $ reflects the topology of the physical space. In the modern
Universe topology changes are strongly suppressed. There exist severe
experimental restrictions which come from oscillation experiments (e.g., see
Ref. \cite{osc} and references therein). This means that now there acts a
superselection rule and the operator $N\left( k\right) $ represents a
constant of motion and may be considered as an ordinary function. However in
the early Universe, during the quantum stage, topology changes should take
place and properties of the function $N\left( k\right) $ are formed then
(e.g., there may appear some excess of field modes \cite{k99}).

Consider as an example the modification of quantum electromagnetic field.
The operator of the electromagnetic interaction is described by the term $ 
V=e\int j^{\mu }A_{\mu }d^{3}x$ (where $e$ is the electron charge and $ 
j^{\mu }$ is the current density of sources). In the extended theory the
number of fields is a variable $A_{\mu }^{a}$ ($a=0,1,...$), one can say
that there exist photons of different ''sorts''. However these sorts are
indistinguishable (there exists the identity principle for fields) and,
therefore, the interaction term should include an additional summation over
the sorts of photons\footnote{ 
The same summation does appear in the total Hamiltonian describing free
photons.} $V=e\sum_{a}\int j^{\mu }A_{\mu }^{a}dx$.

The Green function for photons in the coordinate representation has the form
(e.g., see Ref. \cite{landau}) 
\begin{equation}
D_{\mu \nu }\left( x-x^{\prime }\right) =i\left\langle T\left( A_{\mu
}\left( x\right) A_{\nu }\left( x^{\prime }\right) \right) \right\rangle
\label{10}
\end{equation}
where the averaging out is taken over the vacuum state and the symbol $T$
denotes the chronological ordering. The Fourier transform of this function
in the Feinman gauge is expressed via scalar function $D\left( k\right) $ 
\begin{equation}
D_{\mu \nu }\left( k\right) =g_{\mu \nu }D\left( k\right) .  \label{20}
\end{equation}
In the standard quantum theory in the first approximation $D\left( k\right)
=4\pi /k^{2}=4\pi /\left( \omega ^{2}-{\bf k}^{2}\right) $ (here $k_{\mu
}=\left( \omega ,{\bf k}\right) $). While considering photons of a
particular ''sort'' the same is also valid in the extended theory $ 
D^{a}\left( k\right) =4\pi /k^{2}$. However, it is obvious, that in the
perturbation theory all diagrams will include additional summation over
sorts of photons. Such a summation can be carried out explicitly and,
therefore, it is sufficient to consider a modified Green function 
\begin{equation}
\widetilde{D}\left( k\right) =\sum_{a}D^{a}\left( k\right) =\frac{4\pi
N\left( k\right) }{k^{2}}.  \label{30}
\end{equation}
The nontrivial fact here is that EQFT admits a nontrivial ground state in
which the number of fields $N\left( k\right) $ contains a dependence on the
wave vector $k$ \cite{k99}.

From a formal point of view the expression (\ref{30}) breaks the Lorentz
symmetry. This however, does not mean that the extended theory is not
Lorentz invariant one. The symmetry breaks due to a particular choice of the
field ground state $\Phi _{0}$ which, was shown to possess finite energy and
particle number densities \cite{k99}. Such particles are dark and cannot be
directly observable, but should contribute to the dark matter. Therefore,
the state $\Phi _{0}$ distinguishes a particular prefered system of
coordinates. Since in all internal lines of the diagram technique the
function $\widetilde{D}\left( k\right) $ gets multiplied by $e^{2}$, one can
try to interpreet the modification (\ref{30}) as if a point-like charged
particle acquires a dispersion $e^{2}\left( k\right) =e^{2}N\left( k\right) $ 
. However, such an interpretation should be used somewhat cautiously.
Indeed, this kind of a renoramlization of the charge is valid only for
virtual photons, while in processes involving real photons (external photon
lines) this renormalization does not work.

As it was shown in Ref. \cite{k99} the field ground state $\Phi _{0}$ can be
characterized by occupation numbers of the type 
\begin{equation}
N_{k,n}=\theta \left( \mu -\left( n\omega +\Delta \left( k\right) \right)
\right) ,  \label{40}
\end{equation}
where $N_{k,n}$ is the number of field modes in the quantum state $\left(
k,n\right) $ ($n$ is the number of photons in the given mode), $\theta
\left( x\right) $ is the Heaviside step function, $\mu $ is the chemical
potential, and we added the term $\Delta \left( k\right) $ which is the
minimal energy of the field modes (the vacuum spectral energy density). We
note that the current state of the field theory does not allow to fix the
form of $\Delta \left( k\right) $ and in what follows we shall not specify
it. The standard definition $\Delta \left( k\right) =$ $\frac{1}{2}\omega $
results in the infinite energy density in QFT and requires a
renormalization, while in EQFT this choice produces too small value for the
cutoff \ (see below). However we may expect that $\Delta \left( k\right)
\sim \omega >0$ is an increasing function. Thus, for the mode spectral
density we get 
\begin{equation}
N_{k}=\sum_{n=0}^{\infty }\theta \left( \mu _{k}-n\omega \right) =\left[ 1+ 
\frac{\mu _{k}}{\omega }\right] \,\,,  \label{50}
\end{equation}
where $[x]$ denotes the integer part of the number $x$ and $\mu _{k}=\mu
-\Delta \left( k\right) $. Equation (\ref{50}) shows, in particular, that $ 
N_{k}=0$ as $\mu _{k}<0$ and, therefore, there appears a cutoff $k^{\ast }$
whose value is the solution of the equation $\mu -\Delta \left( k^{\ast
}\right) =0$. \ The standard picture of the electromagnetic field is valid
in the wave number range ($\omega =\omega _{k}=\sqrt{{\bf k}^{2}}$) 
\begin{equation}
\omega _{k}>\mu _{k}>0  \label{60}
\end{equation}
where we have $N_{k}=1$. In what follows, for the sake of simplicity, we set 
$\mu _{k}=\mu $ (thereby neglecting the existence of the field energy of
zero-point fluctuations and of the respective cutoff). Then, (\ref{50})
reads $N_{k}=1+\left[ \mu /\omega \right] $ and in the range $\omega <\mu $
we find the correction to the standard Green function 
\begin{equation}
\widetilde{D}\left( k\right) -D\left( k\right) =\frac{4\pi }{k^{2}}\left[ 
\frac{\mu }{\omega }\right] .  \label{70}
\end{equation}

We note that the consideration above remains also valid in the case of the
linearized gravitational field $h_{\mu \nu }$ (gravitons) with the
replacement of the equation (\ref{20}) by $D_{\mu \nu ,\alpha \beta }\left(
k\right) =\frac{1}{2}\left( g_{\mu \alpha }g_{\nu \beta }+g_{\mu \beta
}g_{\nu \alpha }\right) D\left( k\right) $ and the same function $D\left(
k\right) $ as in (\ref{20}).

Consider now the correction to the Coulomb (or Newton) laws. Since the
number of fields is a variable the Coulomb (or Newton) potential $V$ should
be replaced by the effetive potential $V=\sum_{a}V^{a}$ which contains the
sum over all sorts of photons (gravitons). Consider a rest point-like
particle. Then the Fourier transform of the correction $\delta V\left(
k\right) $ to the standard Coulomb potential $V\left( k\right) =4\pi eZ$ $ 
/\left| {\bf k}\right| ^{2}$ takes the form 
\begin{equation}
\delta V\left( {\bf k}\right) =\frac{4\pi eZ}{\left| {\bf k}\right| ^{2}} 
\left[ \frac{\mu }{\left| {\bf k}\right| }\right] ,  \label{80}
\end{equation}
where $Z$ denotes the value of the electric charge of the rest particle (in
the case of gravity one should use the obvious replacement $eZ\rightarrow -Gm
$). In the coordinate representation this potential is given by the integral 
\begin{equation}
\delta V\left( r\right) =\frac{1}{2\pi ^{2}}\int\limits_{0}^{\infty }\left(
\delta V\left( \omega \right) \omega ^{3}\right) \frac{\sin \left( \omega
r\right) }{\omega r}\frac{d\omega }{\omega }.  \label{90}
\end{equation}
Since $\delta V\left( \omega \right) $ vanishes for $\omega >\mu $, the
upper limit in this integral is $\omega =\mu $. At the low point $\omega =0$
this integral is divergent. However, we note that the interaction between
particles can exist only on scales less than the horizon size $\ell _{h}$.
Thus, as the lowerst limit we must take $\omega \sim 1/\ell _{h}\sim H$
where $H$ is the Hubble constant. This integral can be presented in the form 
\begin{equation}
\delta V\left( r\right) =\sigma \sum\limits_{n=1}^{N^{\ast
}}\int\limits_{H}^{\mu /n}\frac{\sin \left( \omega r\right) }{\omega r} 
d\omega ,
\end{equation}
where $\sigma =2eZ/\pi $ and $N^{\ast }=\mu /H$. In the range $\mu r\ll 1$
this correction produses a constant shift (which gives a finite contribution
to the electronagnetic rest mass of the particle $\delta m\left( eZ\right)
=eZ\delta V$) 
\begin{equation}
\delta V\sim \sigma \mu \sum\limits_{n=1}^{N^{\ast }}\frac{1}{n}\sim \sigma
\mu \ln \left( \frac{\mu }{H}\right) .  \label{100}
\end{equation}
In the opposite asymptotics $\mu r\gg 1$ (but $Hr\ll 1$) we find the
estimate 
\begin{equation}
\delta V\sim -\sigma \mu \ln \left( Hr\right) .  \label{110}
\end{equation}

The expression (\ref{110}) shows essential deviations from the Newtons and
Coulomb laws at scales $r>r_{0}\sim 1/\mu $. In particular, at these scales
the Newton and Coulomb forces acquire the behaviour $1/r$ (instead of $ 
1/r^{2}$). We note that the value $r_{0}$ can be very large and, therefore,
to carry out a direct observation of the correction to the Coulomb potential
is impossible (at macroscopic scales the number of positive and negative
charges is equal and the potential vanishes long before reaching the scale $ 
r_{0}$). However in the case of gravity the situation is different, for the
gravitational potential accumulates and the correction (\ref{110}) has to
leave an imprint in astronomical observations. And indeed, there exists an
indication that such a behaviour really takes place. As is well known
observations show that the leading contribution to the distribution of
matter gives the so-called dark matter which should have an exotic
non-baryonic form and is not directly observable (e.g., see \cite{dm}).
There are several observations which provide evidence for dark matter. One
is measurements of the rotational velocity of galaxies as a function of the
radial distance from the center, the so-called rotation curve (see e.g. \cite
{dm,rc}). According to the standard Newton dynamics the rotation curve of a
disk with an internal mass-distribution that follows the observed brightness
law has to show a Keplerian $r^{-1/2}$ behaviour at large radii. However
measurements \cite{rc} show that $v\left( r\right) =v_{m}$ stays constant,
which implies that the total mass contained within a radius $r$, $M\left(
r\right) $, varies in proportion to $r$ . Indeed, according to the standard
Newtons law the acceleration of a body in a circular orbit of the radius $r$
is $a$ $=$ $GM\left( r\right) /r^{2}$ $=$ $v^{2}\left( r\right) /r$ , which
gives $M\left( r\right) =v_{m}^{2}G^{-1}r$ . This can be interpreted as if
the mass per unit luminosity $M/L$ increases with radius and, therefore, a
large fraction of the total mass of a galaxy is in the form of a
non-luminous, dark component located at large radii. However, if we take
into account the correction (\ref{110}), we find that for $r>r_{0}$ $ 
v^{2}\left( r\right) =v_{m}^{2}\sim 2\mu GM/\pi $ is consistent with the
light distribution ($M/L\sim const$). It is not clear yet which fraction of
the dark matter can be explained by the correction (\ref{110}) to the
Newtons potential (we recall that the ground state $\Phi _{0}$ itself
unavoidably predicts the existence of dark matter \cite{k99}). However, this
allows to give a previous rough estimate for the characteristic scale $
r_{0}\gtrsim 1$ Kpc and so the parameter $\mu $ is really small. To get a
more precise estimate requires the further and more thorough confrontation
with observations.

We note that the idea of a modification of gravity at large scales is not
new, e.g. see  critics of different approaches and the list of references 
in Ref. \cite{SW}. Some approaches use potentials close to (\ref
{110}) as an empirical demand to the modification of the Newton's law (or,
equivalently, introduce an additional force), e.g., see \cite{Sand} and have
no any fundamental theoretical background. On the contrary, in the extended 
quantum field theory the correction (\ref{110}) is inevitable consequence 
of the massless nature of the gravitational field.

\end{document}